\documentclass[iop,apjl]{emulateapj}

\newcommand{\citepeg}[1]{\citep[{e.g.,}][]{#1}}
\newcommand{\kms}{km~s$^{-1}$}
\newcommand{\bh}{SN~2010bh}
\newcommand{\grb}{GRB~100316D}
\newcommand{\bw}{SN~1998bw}
\newcommand{\ajaj}{SN~2006aj}
\newcommand{\dhdh}{SN~2003dh}
\def\swift{{\it Swift}}

\begin{document}

\shorttitle{GRB~100316D / SN~2010bh}
\shortauthors{Chornock et al.}

\title{Spectroscopic Discovery of the Broad-Lined Type Ic Supernova
  2010bh Associated with the Low-Redshift GRB 100316D\altaffilmark{*}}

\author{Ryan Chornock\altaffilmark{1},
Edo Berger\altaffilmark{1},
Emily M. Levesque\altaffilmark{2},
Alicia M. Soderberg\altaffilmark{1,3},
Ryan J. Foley\altaffilmark{1,4},
Derek B. Fox\altaffilmark{5},
Anna Frebel\altaffilmark{1,4},
Joshua D. Simon\altaffilmark{6},
John J. Bochanski\altaffilmark{7},
Peter J. Challis\altaffilmark{1},
Robert P. Kirshner\altaffilmark{1},
Philipp Podsiadlowski\altaffilmark{8},
Katherine Roth\altaffilmark{9},
Robert E. Rutledge\altaffilmark{10},
Brian P. Schmidt\altaffilmark{11},
Scott S. Sheppard\altaffilmark{12},
and Robert~A.~Simcoe\altaffilmark{7,13}
}

\altaffiltext{*}{This Letter includes data gathered with the 6.5-m
  Magellan Telescopes located at Las Campanas Observatory, Chile.}
\altaffiltext{1}{Harvard-Smithsonian Center for Astrophysics, 60
  Garden St., Cambridge, MA 02138, USA,
  \texttt{rchornock@cfa.harvard.edu}.}
\altaffiltext{2}{Institute for Astronomy, University of Hawaii, 2680 Woodlawn Dr., Honolulu, HI 96822, USA .}
\altaffiltext{3}{Hubble Fellow.}
\altaffiltext{4}{Clay Fellow.}
\altaffiltext{5}{Department of Astronomy and Astrophysics, 525 Davey Laboratory, Pennsylvania State University, University Park, PA 16802, USA.}
\altaffiltext{6}{Observatories of the Carnegie Institution of Washington, 813 Santa Barbara Street, Pasadena, CA 91101, USA.}
\altaffiltext{7}{MIT Kavli Institute for Astrophysics \& Space
  Research, Cambridge, MA 02139, USA.}
\altaffiltext{8}{Department of Astrophysics, University of Oxford,
  Oxford OX1 3RH, UK.} 
\altaffiltext{9}{Gemini Observatory, Hilo, HI 96720, USA.}
\altaffiltext{10}{Department of Physics, McGill University, 3600
  University Street, Montreal, QC H3A 2T8, Canada.}
\altaffiltext{11}{Research School of Astronomy and Astrophysics, The
  Australian National University, Weston Creek, ACT 2611, Australia.}
\altaffiltext{12}{Department of Terrestrial Magnetism, Carnegie Institution of Washington, 5241 Broad Branch Rd. NW, Washington, DC 20015, USA.}
\altaffiltext{13}{Alfred P. Sloan Research Fellow.}

\begin{abstract}
We present the spectroscopic discovery of a broad-lined Type Ic
supernova (SN~2010bh) associated with the nearby long-duration
gamma-ray burst (GRB) 100316D.  At $z$ = 0.0593, this is the
third-nearest GRB-SN.  Nightly optical spectra
obtained with the Magellan telescopes during the first week after
explosion reveal the gradual emergence of very broad spectral
features superposed on a blue continuum.  The supernova features are
typical of broad-lined SNe~Ic and are generally consistent with
previous supernovae associated with low-redshift GRBs.  However, the
inferred velocities of \bh\ at 21 days after explosion are a factor of
$\sim2$ times larger than those of the prototypical \bw\ at
similar epochs, with $v \approx$ 26,000~\kms, indicating a larger
explosion energy or a different ejecta structure.  A near-infrared
spectrum taken 13.8 days after explosion shows no strong evidence
for \ion{He}{1} at 1.083~\micron, implying that the progenitor was
largely stripped of its helium envelope.  The host galaxy is
of low luminosity (M$_R \approx -18.5$~mag) and low metallicity ($Z
\lesssim 0.4$ Z$_{\odot}$), similar to the hosts of other low-redshift
GRB-SNe. 
\end{abstract}
\keywords{gamma rays: bursts --- supernovae: individual (SN~2010bh)}

\section{Introduction}

While the connection between long-duration gamma-ray bursts (GRBs) and
Type Ic supernovae (SNe Ic) has been established in broad terms, the
details remain poorly understood \citepeg{wb06}.  Of particular
importance is identifying the physical parameter(s) that distinguish
the $0.1-1\%$ of SNe Ic which give rise to GRBs from those that do
not.  Related to this question is the diversity within the GRB-SN
sample itself.  Progress requires detailed spectroscopic observations, 
which limit in-depth studies of GRB-SNe to the nearest events 
($z\lesssim 0.2$).  The low detection rate of such events, just $\sim 
0.3$ yr$^{-1}$, has hampered observational progress.

Over the past 12 years, detailed spectroscopy has been obtained for
just four GRB-SNe: 980425/1998bw \citep{galama98,patat01}, 030329/2003dh
\citep{hjorth03,stanek03,matheson03}, 031203/2003lw
\citep{malesani04,avishay04}, and 060218/2006aj
\citep{modjaz06,mirabal06,pian06}. 
  In each case, the spectra revealed unusually broad features, 
indicative of large ejecta velocities that are faster than 90-95\% of 
all ordinary SNe Ic \citep{pod04}.  Recently, the broad-lined 
SN\,2009bb \citep{max09} was also shown to produce relativistic 
ejecta typical of GRBs \citep{sod09bb}.

Joining this small set of events, \grb\ was detected as an image
trigger with the \swift\ Burst Alert 
Telescope (BAT) on 2010 March 16.531 UT \citep{gcn10496}.  
The combined pre-trigger BAT survey data and BAT triggered
observations revealed a peculiar GRB with a long duration
($\sim2300$~s), unusual light curve shape, and a soft spectrum 
\citep{gcn10524}.  A slowly-varying X-ray counterpart
was detected with the {\it Swift} X-ray Telescope
(XRT; Starling et al. 2010), but no variable source was
detected with the UV/Optical Telescope in early
observations \citep{gcn10496,gcn10520}.  These peculiarities 
resembled the unusual behavior of the low-redshift XRF 060218
\citep{gcn10511} and indeed a galaxy at $z=0.059$ was found inside the
XRT error circle \citep{gcn10511,gcn10512}.  Follow-up 
observations with the Gemini-South 8-m telescope revealed a
brightening optical counterpart at $\Delta t\approx 1.5-2.5$ d after
the burst, suggestive of an emerging SN \citep{levan_gcn,wiersema_gcn}.

In this Letter, we present our spectroscopic discovery of a SN 
associated with GRB\,100316D \citep{chornockgcn}, subsequently 
designated SN\,2010bh \citep{bufano_cbet,chornockcbet}, and 
describe its spectral evolution using extensive optical/near-IR 
observations spanning $1.47-22$ days after the burst.  As in the case of 
previous GRB-SNe, our spectra reveal that SN\,2010bh is a broad-lined
SN~Ic.  We also find no evidence for helium.  The photospheric
velocity at late time is significantly higher than even SN\,1998bw.
Finally, we show that the metallicity of the explosion 
site is $Z\lesssim 0.4$ Z$_\odot$, in line with the other nearby GRB-SNe
host galaxies.

\section{Observations}

Five observations were performed using the Magellan Echellette (MagE)
spectrograph, while single spectra were obtained using the Low
Dispersion Survey Spectrograph (LDSS3) and the Inamori Magellan Areal
Camera and Spectrograph (IMACS).  Two additional epochs of
spectroscopy were obtained at later times using the Gemini
Multi-Object Spectrograph (GMOS) on Gemini-South.  See
Table~\ref{spectab} for details. 

\begin{figure}
\plotone{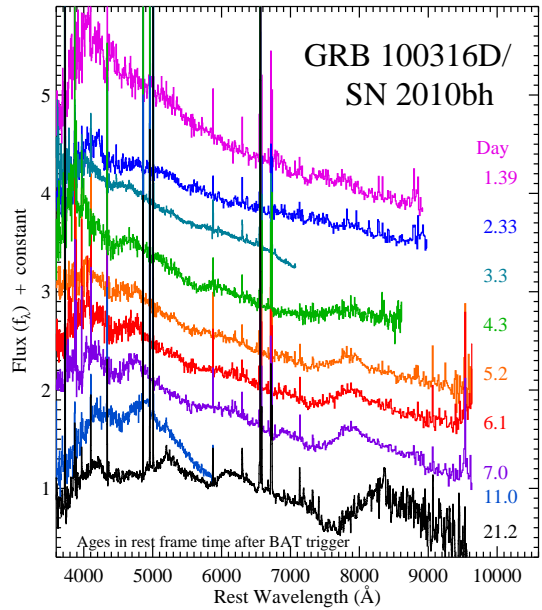}
\caption{Optical spectra of \bh\ from Magellan and Gemini-South
  demonstrating the emergence 
  of the spectral features of a broad-lined SN~Ic.  The spectra
  have been rebinned for clarity.
  The day 11.0 GMOS spectrum has an unreliable overall spectral shape.
}
\label{allplot}
\end{figure}

Standard two-dimensional image reduction and spectral extraction
for the long-slit spectra were performed using IRAF\footnote{IRAF is
  distributed by the National Optical Astronomy Observatories, 
    which are operated by the Association of Universities for Research
    in Astronomy, Inc., under cooperative agreement with the National
    Science Foundation.}.  Flux calibration and removal of telluric
absorption features were performed using our own IDL routines
\citep{ma00}.  The MagE spectra were reduced using custom IDL scripts.
The LDSS3 and IMACS spectra were obtained over a wide wavelength range
without the use of order-blocking filters and the effects
of second-order light contamination are apparent at long wavelengths;
these spectra have been truncated.  The optical spectra are
presented in Figure~\ref{allplot}.  All spectra used in this Letter
have been corrected for Galactic extinction, which is
$E(B-V)=0.116$~mag for \bh\ \citep{sfd98}.  

Since we commenced our spectroscopic observations prior to the
discovery of a variable source, we oriented the MagE slit for the two
initial epochs at a position angle (P.A.) aligned
with the majority of the light from the galaxy.  This strategy was
successful, although it resulted in some contamination from galaxy
light across the entire 10$\arcsec$ slit.  All subsequent observations
were acquired at a more favorable P.A. and clean background
subtraction was obtained.  The first GMOS spectrum was observed at
high airmass with the slit far from the parallactic angle, so its
overall spectral slope is less reliable.

We also obtained a single near-infrared (NIR) spectrum on 2010
March 31.15 during commissioning of the new Folded-port InfraRed
Echellette (FIRE; Simcoe et al. 2008) spectrograph on the Magellan
Baade telescope.  We observed the 0.85$-$2.5~\micron\ range
simultaneously, in the low resolution prism mode.  The resolution
is a strong function of wavelength but is $R = \lambda/\Delta\lambda
\approx 2500$ in the $J$ band.  Two pairs of exposures with
integration times of 600~s and 200~s each were obtained with
\bh\ nodded along the slit.  The OH night sky lines were blended at
this resolution and saturated in $H$ and $K$ in
the longer pair of exposures, so the effective exposure time at
wavelengths longward of 1.4~\micron\ is only 400~s.  The
signal-to-noise ratio in the SN continuum is therefore quite low
in $H$ and $K$, but we can clearly detect narrow
Paschen-$\alpha$ emission from the host galaxy.  We only consider the
spectra at shorter wavelengths hereafter.  Observations of standard
stars were used to apply a flux calibration and correct for telluric
absorption.

\begin{deluxetable*}{cccccccccc}
\tabletypesize{\scriptsize}
\tablecaption{Log of Spectroscopic Observations}
\tablehead{
\colhead{UT Midpoint} & 
\colhead{Age\tablenotemark{a}} & 
\colhead{Instrument} & \colhead{Exp. Time} & 
\colhead{Wavelength} &
\colhead{Slit } & \colhead{Seeing} &
 \colhead{Airmass} &
\colhead{Slit P.A.} &
\colhead{Parallactic Angle} \\
(YYYY-MM-DD.DD)  & (days) & & (s) & (\AA) & (\arcsec) & (\arcsec) & 
 & ($\degr$) & ($\degr$) }
\startdata
2010-03-18.00 & 1.39 & MagE & 1800 & 3300$-$9500 & 1.0 & 0.7 & 1.1 & 100 & $-$4 \\
2010-03-19.00 & 2.33 & MagE & 1800 & 3300$-$10000 & 1.0 & 0.6 & 1.1 & 100 & $-$3 \\
2010-03-20.01 & 3.3 & LDSS3 & 2400 & 3700$-$7500 & 1.0 & 0.6 & 1.1 & 7 & 5 \\
2010-03-21.07 & 4.3 & IMACS & 3600 & 3780$-$9100 & 0.9 & 0.7 & 1.2 & 46 & 41 \\
2010-03-22.00 & 5.2 & MagE & 1800 & 3300$-$10350 & 1.0 & 0.9 & 1.1 & 0 & 2 \\
2010-03-23.00 & 6.1 & MagE & 1800 & 3300$-$10350 & 1.0 & 0.9 & 1.1 & 0 & 0 \\
2010-03-24.00 & 7.0 & MagE & 1800 & 3300$-$10350 & 1.0 & 0.7 & 1.1 & 0 & 3 \\
2010-03-28.18 & 11.0 & GMOS-S & 1200 & 3400$-$6230 & 1.0 & 0.7 & 1.7 & 190 & 87 \\
2010-03-31.15 & 13.8 & FIRE & 1600 & 8500$-$25000 & 0.6 & 0.6 & 1.6 & 10 & 84 \\
2010-04-08.00 & 21.2 & GMOS-S & 1200,1200 & 3400$-$6230,5885$-$10160 & 1.0 & 0.7 & 1.1 & 190 & 32 \\
\enddata
\tablenotetext{a}{In rest frame, relative to the BAT trigger.}
\label{spectab}
\end{deluxetable*}

\section{Results}

Inspection of the two-dimensional spectroscopic frames reveals that
the SN is spatially coincident with bright nebular emission
lines of large equivalent width (also visible in Figure~\ref{allplot}).
Therefore, an \ion{H}{2} region and underlying massive star cluster
contribute flux to our spectra at an unknown level, particularly the
earliest ones when the SN was faint.  Observations at
late times, after the SN has faded, will be necessary to quantify the
contribution of underlying emission to the current dataset, but this
does not affect our conclusions below.  Fits to the nebular emission
lines at the SN location provide a mean redshift of $z = 0.0593$,
which we have adopted for the plots in this paper.

\subsection{SN Spectral Evolution}

The first SN feature present was an absorption dip near 4340~\AA\ on
day 2.33 which strengthened and moved redward to 4460~\AA\ by day
7.0.  On a similar timescale, a brightening optical counterpart was
reported by \citet{wiersema_gcn}.  Combined with the lack of optical
variability at earlier times \citep{gcn10520}, we assume negligible
contamination from an optical afterglow.  Previously, the earliest
reported detection of SN features in a GRB-SN was at 1.88 rest-frame
days in \ajaj\ \citep{mirabal06}, an object which also 
had a relatively weak optical afterglow from its associated GRB.

\begin{figure}
\plotone{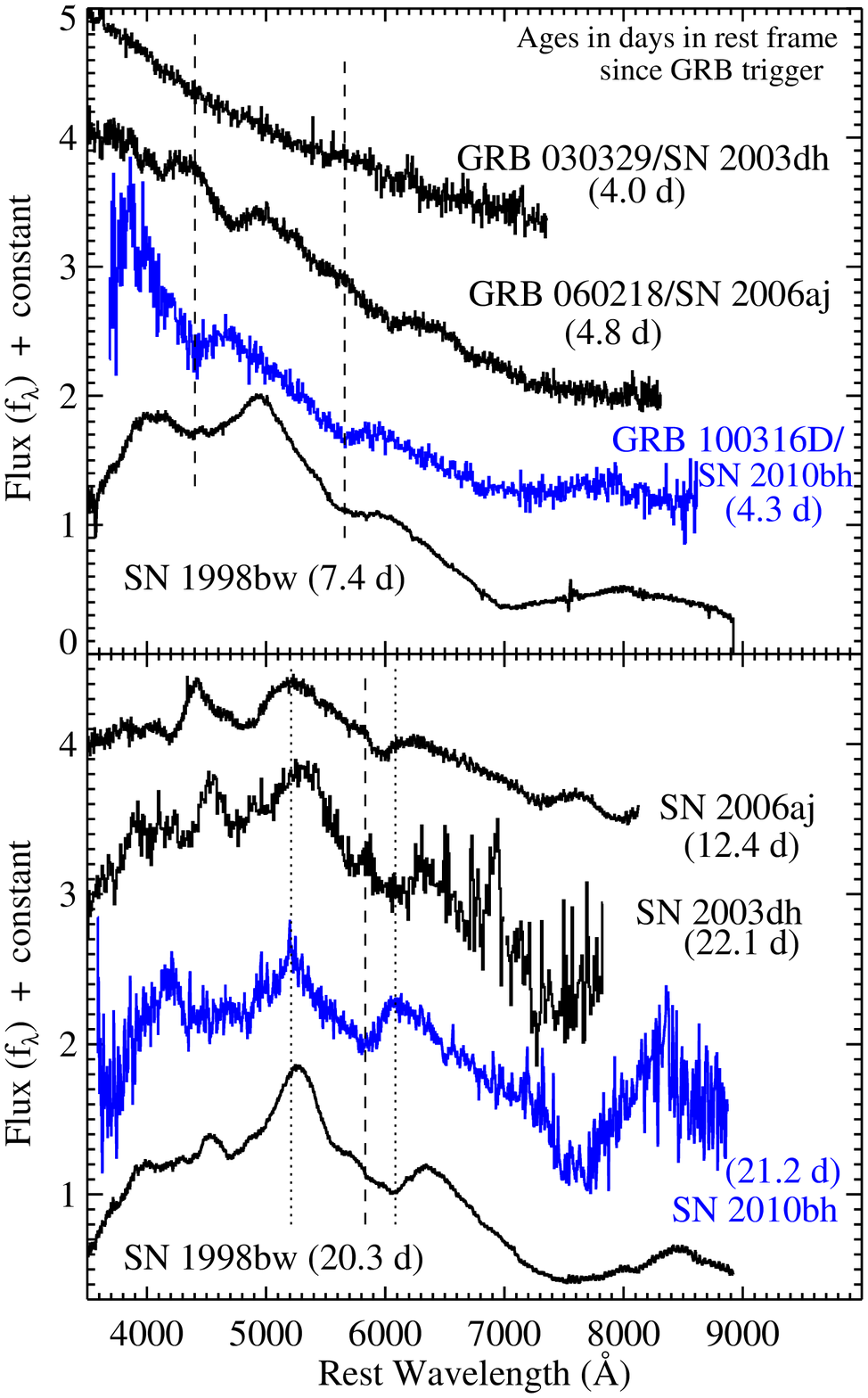}
\caption{Spectral comparison of \bh\ (blue) with other GRB-SNe
  \citep{patat01,matheson03,modjaz06}. 
  Narrow nebular emission lines have been clipped from each spectrum
  for clarity. \emph{Top:} Early-time spectra.  Vertical lines
  (\emph{dashed}) mark the wavelengths of two absorption features in
  \bh.  \emph{Bottom:} Spectra taken near 21 days after explosion.
  The \ion{Si}{2} $\lambda$6355 absorption (\emph{dashed}) and two
  emission peaks (\emph{dotted}) in \bh\ are marked.  All 
  three features are clearly more blueshifted in \bh\ than in the
  others.
}
\label{compplot}
\end{figure}

The spectral features of broad-lined SNe Ic are highly blended.  The
most isolated feature is usually taken to be the minimum near
6000~\AA, which is commonly identified with \ion{Si}{2} $\lambda$6355
\citepeg{patat01}.  That feature is first unambiguously identified
in our data on day 4.3 near 5650~\AA\ (blueshift of 35,000
\kms\ relative to $\lambda$6355), although the day
2.3 data have a very broad undulation with a shallow dip near
5500~\AA\ (43,000 \kms).  A comparison of this spectrum to 
other GRB-SNe is shown in the top panel of Figure~\ref{compplot}.  At
similar epochs, \dhdh\ was still completely dominated by the optical
afterglow of GRB~030329 \citep{matheson03} while the earliest spectrum
of \bw\ is from a few days later but shows stronger spectral
features.  \ajaj\ clearly exhibits lower velocities at this time.

The third emerging SN feature visible in \bh\ on day 4.3 is a broad
undulation with a minimum near 7070~\AA\ and a flux peak near
7840~\AA.  Inspection of Figure~\ref{allplot} shows that corresponding
features are present in the subsequent MagE spectra and smoothly
evolve redward.  By day 7.0, the minimum has moved to 7350~\AA\ and
the peak is near 7880~\AA.  A similar feature is present in the
\bw\ spectrum (Figure~\ref{compplot}), and it was identified as a
blend of \ion{O}{1} $\lambda$7774 and the \ion{Ca}{2} NIR triplet at
high velocities \citep{patat01}.

If we assign the measured absorption minimum in \bh\ to \ion{O}{1}
$\lambda$7774, the implied velocities decrease from 28,300
\kms\ on day 4.3 to 16,700 \kms\ on day 7.0, values which are low
compared to the behavior of the \ion{Si}{2} line.  One interpretation
is that the observed feature is dominated by some other species, such
as the \ion{Mg}{2} $\lambda\lambda$7877, 7896 blend.  If the observed
feature is in fact primarily due to \ion{Mg}{2}, the implied blueshift
at absorption minimum decreases from 32,500 \kms\ on day 4.3 to
21,000 \kms\ on day 7, values which are not only more consistent
with those inferred from the \ion{Si}{2} line, but also lead to a
P-Cygni emission peak near zero velocity.

By the time of our final spectrum, on day 21.2, this feature had
morphed into a strong P-Cygni profile with an emission peak near
8300~\AA\ and an absorption component at 7600~\AA.  These values are
too red for \ion{O}{1} or \ion{Mg}{2} to be contributing significantly
to the feature at this time.  Instead, it appears to be dominated by
the \ion{Ca}{2} NIR triplet.  Relative to the $gf$-weighted line
centroid at 8579~\AA, the absorption minimum is blueshifted by 36,000
\kms.  This is a large velocity, but consistent with the value derived
from \ion{Si}{2} $\lambda$6355 at early times.

\subsection{Large Late-time Velocities}

The \ion{Si}{2} velocity in \bh\ remains large on day 21.2.
This can be seen directly in the bottom panel of
Figure~\ref{compplot}.  The vertical dashed line marks the local flux
minimum in the \bh\ spectrum and it is clearly blueward of the same
feature in \bw\ at a comparable epoch and \ajaj\ at an even younger
age.  Quantitatively, the flux minimum is near 26,000
\kms\ (5820~\AA), while the minimum in 
\bw\ is blueshifted by only 13,000 \kms\ at this epoch.  Similar
conclusions can be drawn from the \ion{Ca}{2} NIR triplet.  As noted
above, in \bh\ at this epoch the flux minimum is blueshifted by 36,000
\kms.  While the \bw\ spectrum is quite blended at these wavelengths
and a broad local minimum is present near 7500~\AA, another
absorption notch is present near 8120~\AA, which corresponds to a
velocity of 16,400 \kms\ relative to \ion{Ca}{2}.  This notch clearly
develops into the \ion{Ca}{2} absorption minimum at later times
\citep{patat01}.

\begin{figure}
\plotone{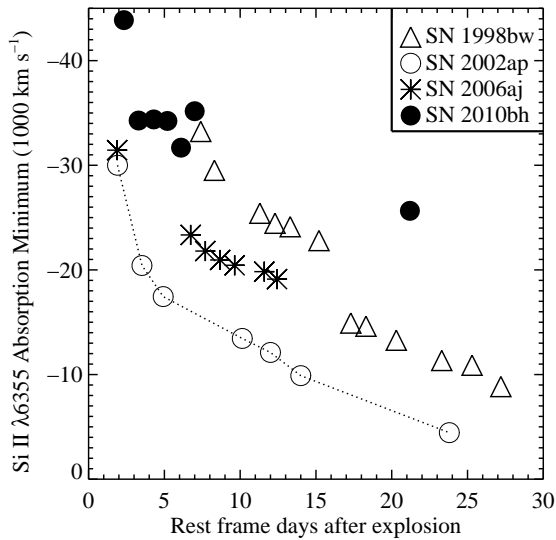}
\caption{Velocity evolution of the absorption minimum near
  5900~\AA\ for several GRB-SNe, assuming that it is \ion{Si}{2}
  $\lambda$6355 in all objects. \bh\ exhibits higher velocities and a
  shallower decline rate than \bw\ at late times, indicative of a 
  higher explosion energy or different ejecta structure.  
  Datapoints for SN~2002ap, which is a broad-lined SN~Ic not
  associated with a GRB, are from the model fits of
  \citet{maz02}.
}
\label{velplot}
\end{figure}

To make a quantitative comparison, we plotted the velocity
evolution of the $\lambda$6355 feature in Figure~\ref{velplot}
compared to \bw\ and \ajaj; the features in \dhdh\ are difficult to
measure against the dilution of the GRB afterglow.  Also plotted are
velocities for the broad-lined SN Ic 2002ap, which had
lower late-time velocities than the GRB-SNe despite the high
early-time velocities.  While we do
not have the same sampling as the \bw\ observations, the rate at which
the velocity  of the \ion{Si}{2} feature decreased with time is lower
in \bh, possibly indicating a different ejecta structure or a more
powerful explosion, although complications such as potential
differences in the ejecta mass preclude any firm statements.

\bh\ also differs from \dhdh\ and \ajaj\ (while resembling \bw) in its
lack of an emission bump near 4500~\AA\ (Figure~\ref{compplot}).  This
feature has also been seen in broad-lined SNe Ic not associated with
GRBs, such as SN~1997ef \citep{iwamoto00} and SN~2002ap
\citep{foley03}.  Even normal-velocity SNe~Ic such as SN~1994I
\citep{fil95} have an emission peak near these 
wavelengths.  Due to the highly blended nature of the spectral
features in broad-lined SNe Ic, the emission peaks represent minima in
the line opacity rather than emission features from a single
transition \citepeg{iwamoto00}.  The lack of the 4500~\AA\ feature in
\bh\ at an epoch where it is present in other SNe Ic may therefore be
another consequence of the unusually high velocities at late times if
the iron lines to the blue and red of this wavelength are still
blended together.  Similarly, the emission peaks present in \bh\ near
5200~\AA\ and 6085~\AA\ on day 21.2 are more blueshifted than in the
other objects at that time.

Two caveats for this analysis are that the \ion{Si}{2} line could be
blended with other features in a manner that compromises our velocity
measurements or the \ion{Si}{2} line-formation layer could be
unrepresentative of the true photospheric velocity at late times.
However, we believe the combined evidence favors significantly higher
velocities in \bh\ than \bw\ at late times.  We note that effects of
SN explosion asphericity \citep{maeda06} also could complicate the
interpretation of these velocity measurements, but that nebular-phase
spectroscopy will offer some insight on the explosion geometry
\citepeg{maz05}. 

\subsection{NIR Spectrum and Lack of Evidence for He}

Our goal in obtaining the FIRE spectrum was to test for the
existence of helium in \bh.  The presence or absence of
helium is a significant constraint on the nature and evolutionary
state of GRB progenitors prior to the explosion.  Absence of helium
is difficult to deduce from optical spectra alone, both due to the
high degree of line blending and the general need for non-thermal
excitation of the optical lines \citepeg{lucy91}.  Both of these
concerns are mitigated by searching for the lower-excitation
\ion{He}{1} lines at 1.083 and 2.056~\micron\ in the less-blended NIR
part of the spectrum.  Furthermore, detection of these NIR lines is
substantially more difficult for GRB-SNe at high redshift, so nearby
objects represent our best chance to constrain the presence of helium
in GRB progenitors.  \citet{patat01} claimed that both of these lines
were present in \bw, although the weakness of the observed
2.056~\micron\ line has made this result controversial. 

\begin{figure}
\plotone{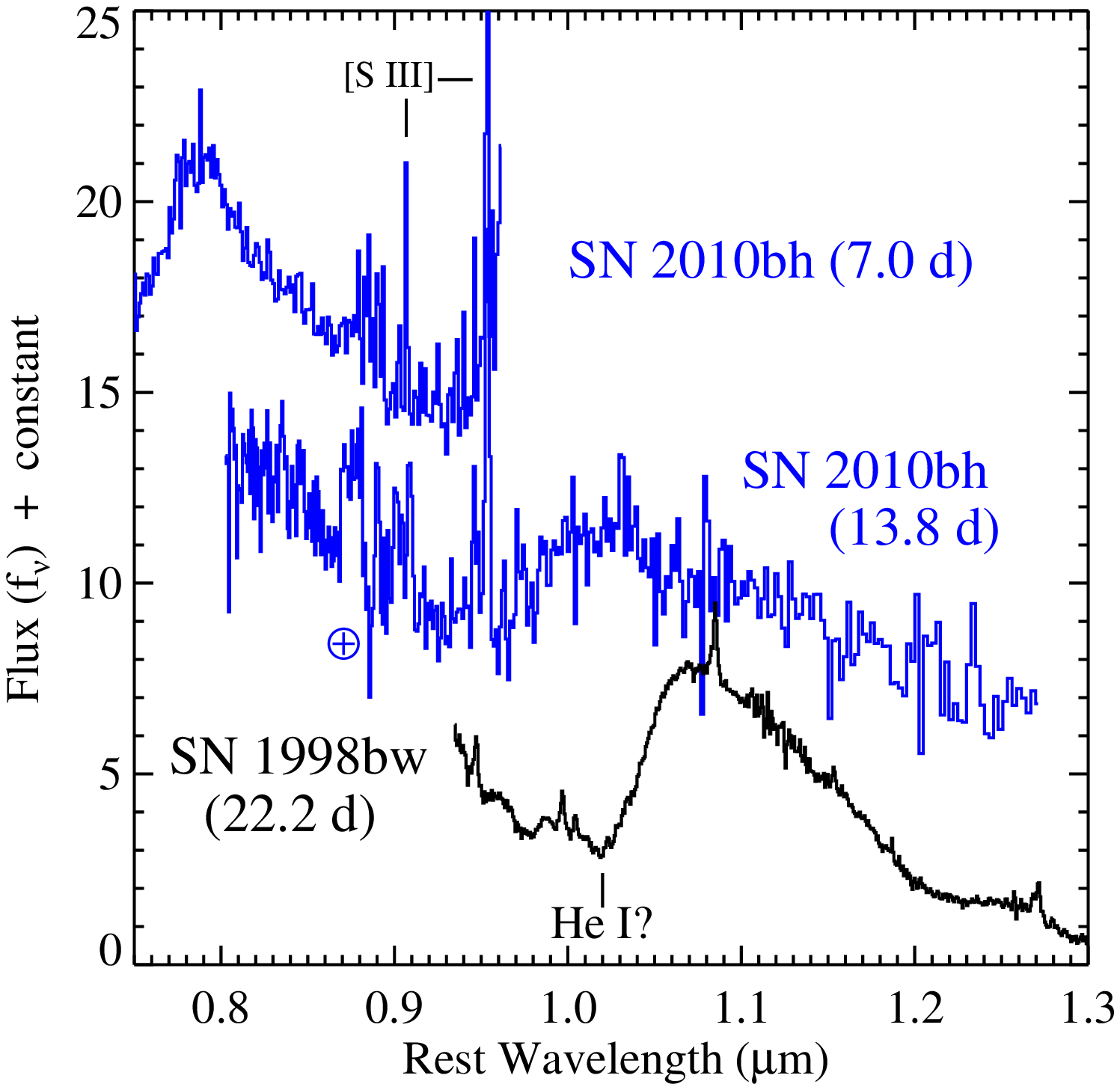}
\caption{FIRE spectrum of \bh\ (middle) compared to the
  long-wavelength end of the day 7.0 MagE
  spectrum (top) and the earliest NIR spectrum of
  \bw\ (bottom; Patat et al. 2001).  The \bw\ spectrum has an
  absorption minimum near
  1.02~\micron\ which has been identified with \ion{He}{1} 1.0830
  \micron, but no clear corresponding feature is present in \bh.
  A poorly-removed telluric feature in the \bh\ spectra is marked with
  a $\earth$ symbol.  Two narrow [\ion{S}{3}] emission lines from the
  host galaxy of \bh\ are marked.
}
\label{nirplot}
\end{figure}

Our FIRE spectrum is shown in Figure~\ref{nirplot} and includes the
wavelength region around the 1.083~\micron\ line of \ion{He}{1}.  A
broad spectral feature is present with a peak near 1.02~\micron\ and
an absorption minimum to the blue near 0.95 \micron.  We have also
plotted the earliest \bw\ NIR spectrum, with the absorption minimum at
1.02~\micron\ identified as \ion{He}{1} by \citet{patat01} labeled.
No corresponding feature is present in
\bh.  If the feature at 0.95~\micron\ were identified with
\ion{He}{1} absorption, the inferred velocity would be nearly 40,000
\kms, much higher than the photospheric velocity measured in the
optical at that time.  We conclude that there is no strong evidence
for helium in the outer ejecta of \bh\ on day 13.8, unless it is
present at extremely high velocities. 

The identity of the absorption near 0.95 \micron\ is unknown.  SNe
Ib/c frequently have an absorption present redward of 1 \micron\ and
several candidate ions have been suggested to contribute to that
feature, such as \ion{Si}{1} and \ion{C}{1} \citep{gerardy04,taub06},
although \ion{C}{1} would be expected to exhibit other, weaker
transitions as well \citepeg{sauer06}.  If the absorption velocity of
the feature is similar to that measured for \ion{Si}{2} in the
optical, the implied rest wavelength is 1.04$-$1.05 \micron.

\subsection{Host Galaxy Metallicity}

We used the LDSS3 spectrum from day 3.3 to extract information on the
bright star-forming region underlying \bh.  The observed ratio of
$F_{\rm H\alpha}/F_{\rm H\beta}\approx 2.82$ is close to the Case B
recombination value, 
from which we infer that the reddening of the \ion{H}{2} region is
negligible.  The blue spectral shape of the early spectra is an
additional argument that \bh\ itself suffers low extinction.  We
estimate an age for the young stellar population of about 6 Myr from
the H$\beta$ equivalent width ($EW\approx22.8$~\AA; Levesque et
al. 2010), although this is an upper limit since the SN continuum
dilutes the measured EW relative to its true value.

We detect the auroral [\ion{O}{3}] $\lambda4363$ emission line in our
spectra.  This is often a good indicator of low metallicity because
the weakness of the line renders it unobservable in higher-metallicity
environments and, indeed, we directly measure ${\rm
  log(O/H)}+12\approx 7.8$ from the derived electron temperature,
$T_e$.  
For comparison with studies of previous GRB-SN and long GRB host
galaxies (e.g., \citealt{lbkb10}), we also derived the metallicity
from strong-line diagnostics.  Our $T_e$ results place the
\bh\ environment on the lower branch of the $R_{23}$ metallicity
diagnostic \citep{kd02,kk04}, which yields a metallicity of ${\rm
  log(O/H)}+12\approx 8.3$.  In addition, we also calculate a
metallicity of ${\rm log(O/H)}+12 \approx 8.2$ based on the O3N2
diagnostic \citep{pp04}\footnotemark\footnotetext{The $T_e$-based
  methods are known to yield systematically lower metallicities than
  those determined from strong-line diagnostics \citep{kbg03,ke08}.
  The metallicities determined here are all consistent with these
  offsets.}.
The strong-line diagnostics give $Z \lesssim 0.4$ Z$_{\odot}$, which is
similar to the low-metallicity environments of previous GRB-SNe when
placed on the same abundance scales.  Additionally, we
estimate a total magnitude for the host galaxy of M$_R \approx
-18.5$~mag from our IMACS acquisition images, corresponding to $\sim
0.1$~L$_*$, with some uncertainty due to correction for the emission
from the SN.

\section{Conclusions}
The spectra presented in this Letter provide conclusive evidence that
the $z=0.0593$ GRB 100316D was accompanied by the broad-lined SN~Ic
2010bh.  We highlight the following basic properties:

\begin{enumerate}

\item Spectral features from \bh\ were detected beginning 2.33 days
  after the GRB, one of the earliest such detections in a GRB-SN.

\item The \ion{Si}{2} velocities on day 21 were twice as large as
  \bw.

\item The velocity evolution was slower in \bh\ than in
  other GRB-SNe, possibly indicating diversity in the ejecta structure
  of these events.

\item We find no evidence for helium, indicating a highly-stripped
  progenitor star.

\item The SN is superposed on a star-forming region with a
  low-metallicity ($Z \lesssim 0.4 Z_{\odot}$) in a low-luminosity
  host galaxy.

\end{enumerate}

Continued observations to the nebular phase will shed further light on
the ejecta velocity and geometry.

\acknowledgments
We thank Las Campanas Observatory and the CfA and Carnegie Time
Allocation Committees for supporting the Target of Opportunity
interrupt observations which made this study possible.  We acknowledge
the staffs at Magellan and Gemini-South for their assistance with
these observations.  The MagE reduction routines were modifications of
the \texttt{mage\_reduce.pro} IDL scripts written by G. D. Becker. 
We acknowledge support from NASA/Swift Guest Investigator grant
NNX09AO98G.  
Some observations were obtained at the Gemini Observatory (Program
ID: GS-2010A-Q-5),
which is operated by the Association of Universities for Research in
Astronomy, Inc., under a cooperative agreement with the NSF on behalf
of the Gemini partnership: the National Science Foundation (United
States), the Science and Technology Facilities Council (United
Kingdom), the National Research Council (Canada), CONICYT (Chile), the
Australian Research Council (Australia), Minist\'{e}rio da Ciencia e
Tecnologia (Brazil) and Ministerio de Ciencia, Tecnolog\`{i}a e
Innovaci\'{o}n Productiva (Argentina).

{\it Facilities:} \facility{Magellan:Baade (IMACS, FIRE)},
\facility{Magellan:Clay (MagE, LDSS3)}, \facility{Gemini:South
  (GMOS-S)}

\end{document}